\documentclass[%
 aip,
 amsmath,amssymb,
 reprint,%
 widetext
]{revtex4-1}

\usepackage{svg}
\usepackage{amsmath}
\usepackage{graphicx}
\usepackage{dcolumn}
\usepackage{bm}

\usepackage[utf8]{inputenc}
\usepackage[T1]{fontenc}
\usepackage{mathptmx}
\usepackage{etoolbox}
\usepackage{soul}
\usepackage{xcolor}

\DeclareMathOperator{\sgn}{sgn}
\DeclareMathOperator{\E}{\mathbf{E}}
\DeclareMathOperator{\abs}{abs}
\DeclareMathOperator{\ind}{\mathbf{1}}

\makeatletter
\def\@email#1#2{%
 \endgroup
 \patchcmd{\titleblock@produce}
  {\frontmatter@RRAPformat}
  {\frontmatter@RRAPformat{\produce@RRAP{*#1\href{mailto:#2}{#2}}}\frontmatter@RRAPformat}
  {}{}
}%
\makeatother

\begin{document}

\title{Solving the Wigner Equation for Chemically Relevant Scenarios: Dynamics in 2D}

\author{Yu Wang}
\author{Lena Simine}%
    \email{lena.simine@mcgill.ca}
    \affiliation{Department of Chemistry, McGill University, 801 Sherbrooke West, Montreal, QC, Canada}%

\date{yesterday}

\begin{abstract}

Signed Particle Monte Carlo (SPMC) approach has been used in the past to model steady-state and transient dynamics of the Wigner quasi-distribution for electrons in low dimensional semiconductors. Here we make a step towards high-dimensional quantum phase-space simulation in chemically relevant scenarios by improving the stability and memory demands of SPMC in 2D. We do so by using an unbiased propagator for SPMC to improve trajectory stability and by applying machine learning to reduce memory demands for storage and manipulation of the Wigner potential. We perform computational experiments on a 2D double-well toymodel of proton transfer and demonstrate stable pico-second-long trajectories that require only a modest computational effort.

\end{abstract}

\maketitle

\section{Introduction}

The Wigner function was proposed when forerunners of quantum mechanics were seeking quantum corrections to thermodynamic equilibrium \cite{wigner32}. The result was a new more intuitive formulation of quantum mechanics analogous to classical mechanics because the dynamics took place in the phase space. The equation of motion for the Wigner quasi-distribution function, a partial integro-differential equation, was rather difficult to solve \cite{wigner84, tannor97,ferry} and phase-space formulation attracted faltering attention over the years. Nonetheless, important contributions were made to simulations of quantum and semi-classical dynamics in the fields of quantum optics, quantum transport, molecular dynamics, and quantum information - for an extensive review see Ref. \cite{ferry18} and the references therein.

In spite of the inherent difficulties, recently there was a resurgence of interest in the Wigner equation\cite{ferry18,prb59,prb88} in the context of simulation of quantum transport in semiconductor devices \cite{rmp83, rmp83a, rmp90, rmp91}. This is due to the fact that the Wigner equation can express open quantum systems naturally while retaining similarities with Boltzmann transport equations. New algorithmic developments in this space, including the algorithm that is the focus of our paper, provide new pathways to investigation of semiconducting materials using the Wigner equation \cite{pat}. 

Moving beyond charge dynamics in semiconductors we seek to bring this novel computational approach\cite{muscato, jm14, jm15, jm15a, jm15ho, jmbeijing15} into chemical physics. Previously, we have explored the stability of the solutions produced by the Signed Particles Monte Carlo method (SPMC) for electronic and for nuclear dynamics in simple 1D potentials which were parameterized to match the typical energy scales encountered in chemical scenarios \cite{ours}. Here we advance our simulations into 2D using a toy model double well potential that was developed for modelling a hydrogen transfer reaction in salicylaldehyde\cite{makri87, rom91}.
 
 In order to make the simulation more stable we update the form of the propagator to improve accuracy of the Monte Carlo protocol. One of the difficulties that one encounters in simulations of chemical systems using the Wigner equation is that polynomial potentials commonly employed in chemical modeling give rise to Wigner potential singularities which may lead to sudden 'explosions' of signed particles in the system within an infinitesimal time step making simulations unstable. Since chemical trajectories tend to be low energy and chemical potentials tend to be anharmonic, here we truncate the pathological polynomial potential at sufficiently high energy to ensure that truncation does not affect low energy dynamics, and use a minimal feed-forward single-layer perceptron model with Gaussian activation functions to fit the truncated potential\cite{mlmit,dl,pytorch}. The outcome can be easily Wigner-transformed to yield an analytical expression for the Wigner potential which takes very little memory to store and results in a light and robust implementation of the SPMC algorithm. This sets the stage for attempting higher-dimensional simulations of quantum dynamics in the phase space in the future.

\section{Signed Particle Monte Carlo (SPMC)}

\subsection{Equation of Motion}

The Wigner function of a quantum system is given by the Wigner transformation of the density operator with respect to relative coordinates
\begin{equation}\label{def}
	f_w(\mathbf{x}, \mathbf{p}) \equiv \frac{1}{{(2\pi\hbar)}^n}
	\int_{\Bbb{R}^n} \left\langle \mathbf{x} - \frac{\mathbf{s}}{2}
	|\ \rho(\mathbf{x})\ | \mathbf{x} + \frac{\mathbf{s}}{2} \right\rangle\
	e^{i\mathbf{p} \cdot \mathbf{s}/\hbar}\ d^n\mathbf{s}.
\end{equation}

\noindent
The time evolution of the Wigner equation is found by substituting the definition \eqref{def} into the Liouville--von Neumann equation
\begin{equation}\label{eom}
	\frac{\partial f_w}{\partial t} + \frac{\mathbf{p}}{m} \cdot
	\nabla_x f_w = \int_{\Bbb{R}^n} V_w(\mathbf{x}, \mathbf{p} - \mathbf{p}')
	f_w(\mathbf{x}, \mathbf{p}')\ d^n\mathbf{p}'.
\end{equation}

\noindent
Here $V_w$ is the Wigner potential and is defined by the Fourier transform of the central difference of the potential $V(x)$
\begin{equation}\label{vw}
	V_w(\mathbf{x}, \mathbf{p}) \equiv \frac{1}{{i\hbar (2\pi\hbar)}^n}
	\int_{\Bbb{R}^n} \left[ V\Big( \mathbf{x} + \frac{\mathbf{s}}{2} \Big) -
	V \Big( \mathbf{x} - \frac{\mathbf{s}}{2} \Big) \right]
	e^{-i\mathbf{p} \cdot \mathbf{s}/\hbar}\ d^n\mathbf{s}.
\end{equation}

\noindent
The equation of motion \eqref{eom} is known as the Wigner equation. It reduces to the Boltzmann transport equation in the classical limit or for quadratic potentials. The scattering integral is the convolution of the Wigner potential and the Wigner function. If we compare it to the classical Boltzmann transport equation, we see that it is similarly characterized by Newtonian trajectories but with the acceleration term missing on the left-hand side of Eq.\ \eqref{eom},
\begin{equation}\label{traj}
	\mathbf{x}(t) = \mathbf{x}_0 + \frac{\mathbf{p}t}{m},
\end{equation}

\noindent
where $m$ is the mass of the particle. We may rewrite the Wigner potential in Eq.\ \eqref{vw} as a sum of the positive and the negative parts by defining 
\begin{equation}\label{vwpdef}
    V_w^\pm(x,p) \equiv \max \left\{ \pm V_w(x,p),\ 0 \right\}.
\end{equation}

\noindent
Since $V_w(x,p)$ is anti-symmetric with respect to momentum $p$, we may replace $V_w^-(x,p)$ with $V_w^+(x,-p)$ and arrive at the form that will be shown to be convenient for Monte Carlo integration,
\begin{equation}\label{vwplus}
	V_w(\mathbf{x}, \mathbf{p}) = V_w^+(\mathbf{x}, \mathbf{p}) - V_w^-(\mathbf{x}, \mathbf{p}) 
	= V_w^+(\mathbf{x}, \mathbf{p}) - V_w^+(\mathbf{x}, -\mathbf{p}).
\end{equation}

\noindent
This decomposition separates the Wigner potential into two positive components which may be normalized and interpreted as probabilities.

\subsection{Short-time Propagator of the Wigner Equation}\label{sectionB}

Using the commutative property of the convolution, Eq.\ \eqref{eom} can be cast into the following form:
\begin{multline}\label{eom2}
	\frac{\partial f_w}{\partial t} + \frac{\mathbf{p}}{m} \cdot
	\nabla_x f_w = \frac{1}{2} \int_{\Bbb{R}^n}
	V_w(\mathbf{x}, \mathbf{p}') \left(
	f_w(\mathbf{x}, \mathbf{p} - \mathbf{p}') \right.\\
	- \left. f_w(\mathbf{x}, \mathbf{p} + \mathbf{p}')
	\right) d^n\mathbf{p}'.
\end{multline}

\noindent
We define an auxiliary function $\gamma(x)$
\begin{equation}\label{gama}
	\gamma(\mathbf{x}) \equiv \frac{1}{2} \int_{{\Bbb{R}}^n} V_w^+(\mathbf{x}, \mathbf{p})\ d^n\mathbf{p},
\end{equation}

\noindent
and then add the term $\gamma(\mathbf{x}) f_w(\mathbf{x}, \mathbf{p})$ to both sides of Eq.\ \eqref{eom} and obtain

\begin{widetext}

\begin{equation}\label{eomg}
	\frac{\partial f_w}{\partial t} + \frac{\mathbf{p}}{m} \cdot
	\nabla_x f_w + \gamma(\mathbf{x}) f_w
	= \int_{\Bbb{R}^n} \frac{V_w(\mathbf{x}, \mathbf{p}')}{2}
	\left( f_w(\mathbf{x}, \mathbf{p} - \mathbf{p}') 
	- f_w(\mathbf{x}, \mathbf{p} + \mathbf{p}') \right)
	+ \gamma f_w(\mathbf{x}, \mathbf{p}')
	\delta(\mathbf{p} - \mathbf{p}')\ d^n\mathbf{p}'.
\end{equation}

\noindent
We parameterize Eq.\ \eqref{eomg} by the trajectory Eq.\ \eqref{traj} and express the Liouville operator in terms of a total derivative of time
\begin{equation}\label{ddt}
	\frac{d}{dt} = \frac{\partial}{\partial t} + \frac{\mathbf{p}}{m} \cdot \nabla_x.
\end{equation}

\noindent
Making use of Eq.\ \eqref{ddt}, the equation of motion can be cast into
\begin{gather}\label{eomt}
	\frac{d f_w(\eta_t)}{dt} + \gamma_t f_w(\eta_t)
	= \int_{\Bbb{R}^n} \frac{V_w(\eta_t')}{2}
	\left( f_w(\eta_t^\Delta) - f_w(\eta_t^\Sigma) \right)
	+ \gamma_t f_w(\eta_t') \delta(\mathbf{p} - \mathbf{p}')\ d^n\mathbf{p}' \\
    \gamma_t = \gamma(\mathbf{x}(t)) \qquad
    \eta_t = (\mathbf{x}(t), \mathbf{p}) \qquad
    \eta_t' = (\mathbf{x}(t), \mathbf{p}') \qquad
    \eta_t^\Delta = (\mathbf{x}(t), \mathbf{p} - \mathbf{p}') \qquad
    \eta_t^\Sigma = (\mathbf{x}(t), \mathbf{p} + \mathbf{p}').
    \notag
\end{gather}

\noindent
We integrate Eq.\ \eqref{eomt} through a small time step
\begin{equation}\label{eomdt}
	f_w(\eta_{t + dt}) = (1 - \gamma_t dt) f_w(\eta_t) + \gamma_t dt
	\int_{\Bbb{R}^n} \frac{V_w(\eta_t')}{2\gamma_t}
	\left( f_w(\eta_t^\Delta) - f_w(\eta_t^\Sigma) \right)
	+ f_w(\eta_t') \delta(\mathbf{p} - \mathbf{p}')\ d^n\mathbf{p}'.
\end{equation}

\end{widetext}

\noindent
The right-hand side of this equation is the small time propagator which we use to evolve the system in time. The propagator in Eq.\ \eqref{eomdt} is an improvement to previous studies.\cite{jm14,jm15} Fluctuations are reduced by splitting $f_w$ into two terms, flip the sign of one term, and then add them back together, which is a common technique.

\subsection{The Signed Particles Monte Carlo Protocol}\label{2c}

Here we provide a brief account of the SPMC simulation protocol, for a similar summary and a step-by-step pseudo-code see Ref.\ \cite{ours,muswag16}. To run Monte Carlo simulations, we first discretize the Wigner function into an ensemble of volume elements as in hydrodynamics.\cite{fluid} The discretization process can be cast into a pictorial view in which the Wigner function is decomposed into a sum of impulses at strategically chosen points in the phase space:
\begin{equation}\label{delta_particles}
    f_w(\eta) \approx \frac{1}{n_+ - n_-} \sum_i
    \sgn\big( f_w(\tilde{\eta}_i) \big)\ \delta(\eta - \tilde{\eta}_i),
\end{equation}

\noindent
where $\tilde{\eta}_i = (\mathbf{x}_i, \mathbf{p}_i)$ is the phase space point where the $i$th impulse located. $n_+$ and $n_-$ refers to the normalization condition $\int f_w\ d\eta = 1$, and is defined as
\begin{equation}
    n_\pm = \sum_i \sgn \big( f_w(\eta_i) \big),
    \quad \text{if} \; f_w(\eta_i) \gtrless 0
\end{equation}

\noindent
We define
\begin{equation}
    u_i = \frac{\sgn\big( f_w(\tilde{\eta}_i) \big)}{n_+ - n_-}
\end{equation}

\noindent
to be the weighted sign. The delta function impulses along with the preceding sign $u_i$ are what we called the ``singed particles'', which evolve ballistically according to Eq.\ \eqref{traj}. We use Monte Carlo integration to evaluate the scattering integral in Eq.\ \eqref{eomdt} which is transformed into

\begin{widetext}

\begin{equation}\label{deltas}
    f_w(\eta_{t+dt}) = \E \Bigg[ \sum_{i=1}^N u_i \delta(\eta_t
    - \tilde{\eta}_t^i) + \gamma_tdt \sum_{i,j = 1}^{N,M}
    \sgn\bigg( \frac{V_w(\tilde{\eta}_t^j)}{2\gamma} \bigg)
    \Big( (-u_j) \delta \big( \eta_t - (\tilde{\eta}_t^i
    - \tilde{\eta}_t^j) \big) + u_j \delta \big( \eta_t
    - (\tilde{\eta}_t^i + \tilde{\eta}_t^j) \big) \Big) \Bigg],
\end{equation}

\end{widetext}

\noindent
where $M$ is the number of samples taken to estimate the integral, and is the total number of scattering events occurred. We see in Eq.\ \eqref{deltas} that at each step old particles remain and new signed particles are added according to the probability $\gamma_tdt$ by the scattering kernel. In appendix A, we derive Eq.\ \eqref{deltas} and show that it is an unbiased estimator to the Wigner function.

For each existing particle that is being scattered with probability $\gamma_tdt$, a pair of +/- particles is added with the shifted momenta coordinates $\tilde{\eta}_t = \tilde{\eta}_t^i \pm \tilde{\eta}_t^j$. The momentum shift $\eta'$ of the new particles is drawn from the probability density $\abs( V_w/2\gamma )$, with the sign of the particles assigned by $\pm \sgn( V_w/2\gamma ) \sgn( f_w(\eta_t) )$. The normalized quasi-probability distribution $f_w(\eta_{t+dt})$ is obtained by adding all terms together with the net outcome of (positive) particles and (negative) antiparticles on the same grid point annihilating each other thereby removing excess particles. The particle-antiparticle balance at any given grid point determines the magnitude of $f_w(\eta_t)$. This value can be positive or negative reflecting the fact that the Wigner function is a quasi-probability function that may take negative values, which is a consequence of the quantum uncertainty principle. In contrast to other approaches, the density of particles does not represent the absolute value of the Wigner function with the sign stored separately, but rather the sign of the Wigner function comes from the fact that particles are signed.

\begin{figure}[b]
    \includegraphics[width=\columnwidth]{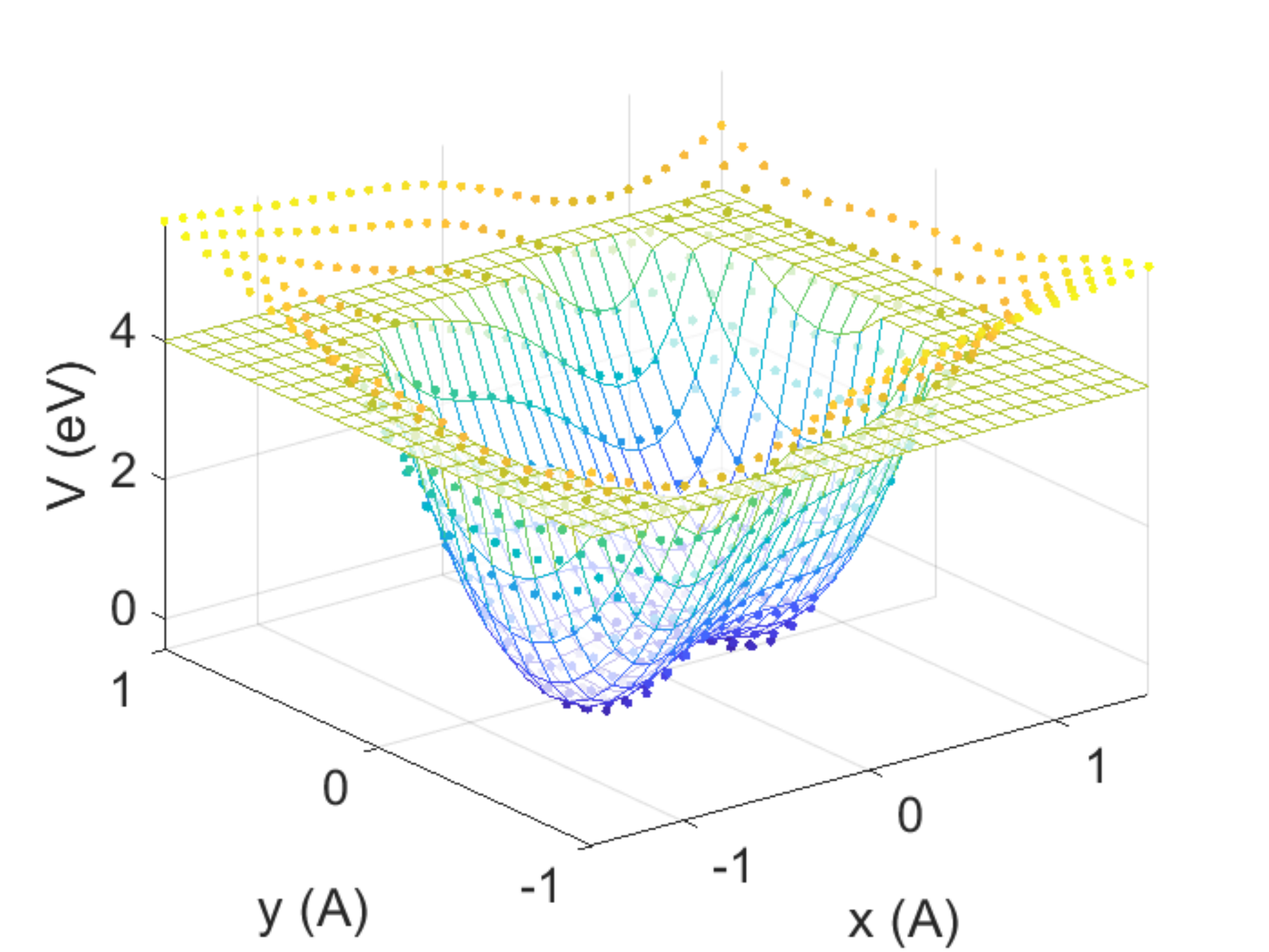}
    \caption{A 2D double well potential describing a proton transfer reaction is shown (mesh). The potential is truncated at an energy sufficiently high to minimally affect the low energy dynamics of the system (energy threshold is set to 4eV). A fitted potential with a Gaussian Neural Network is superimposed (dots) to demonstrate agreement. The behavior of the fitted potential above the truncation threshold is assumed to be inconsequential.}
    \label{fitted}
\end{figure}

\subsection{Handling Singularities in Wigner Potentials}\label{2d}

The auxiliary function $\gamma$, defined in Eq. \eqref{gama}, is physically interpreted as the rate of creating signed particles in the Monte Carlo simulation. For polynomial potentials which appear often in chemically-relevant scenarios $V_w^+$'s are derivatives of delta functions and this results in unbounded rate of signed particles creation. Furthermore, in this case, it is not possible to separate $V_w$ into $V_w^+$ and $V_w^-$ from the perspective of the Cauchy principle value because both of them are unbounded. This is the singularity of the scattering integral. From Eqs.\ \eqref{gama} and \eqref{eomdt} we see that unbounded $V_w^+$ requires a vanishing time step in order to keep the product $\gamma dt$ finite rendering the simulation hopelessly infeasible.

\begin{figure*}
    \includegraphics[width=\columnwidth]{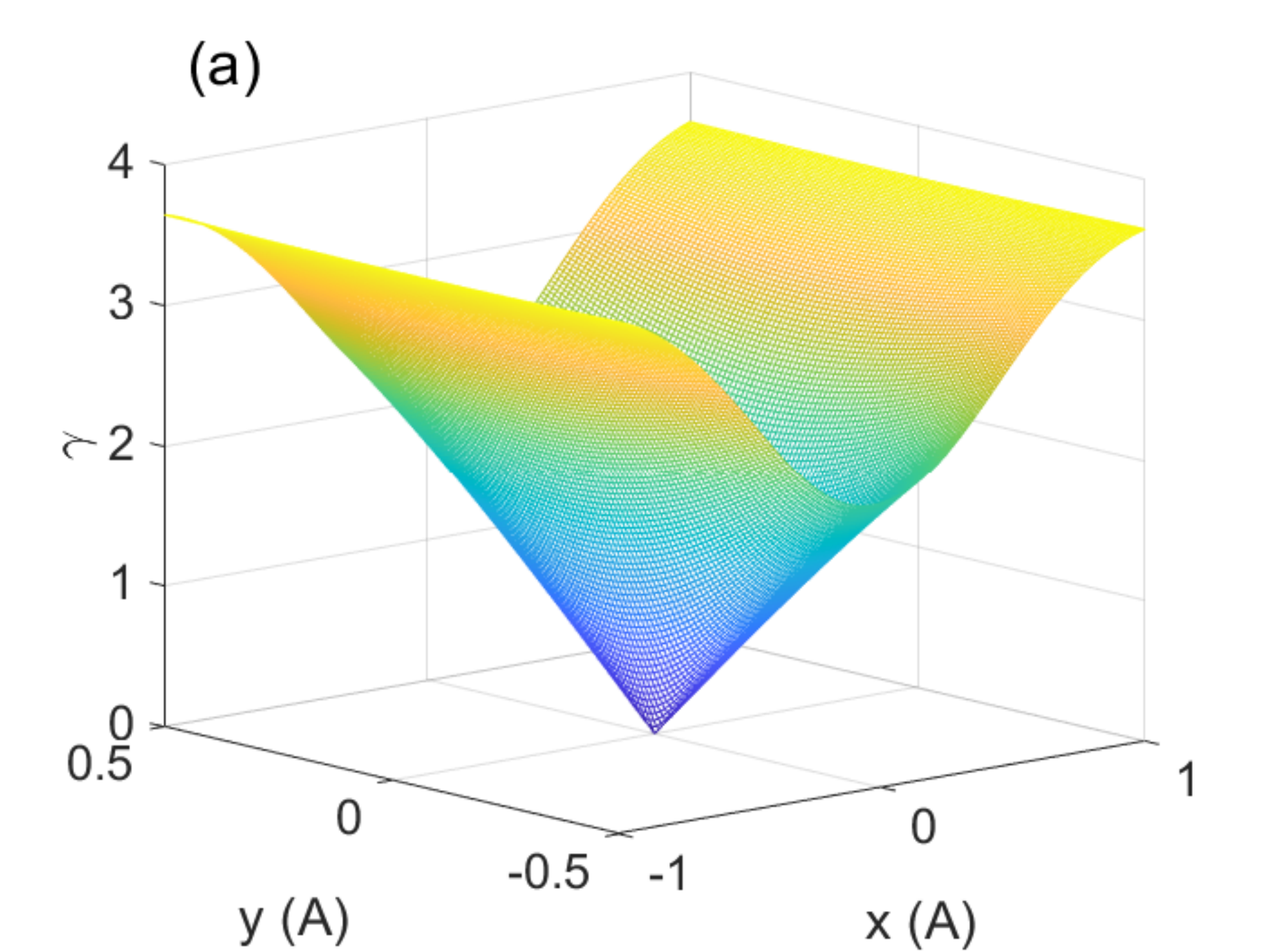}%
    \includegraphics[width=\columnwidth]{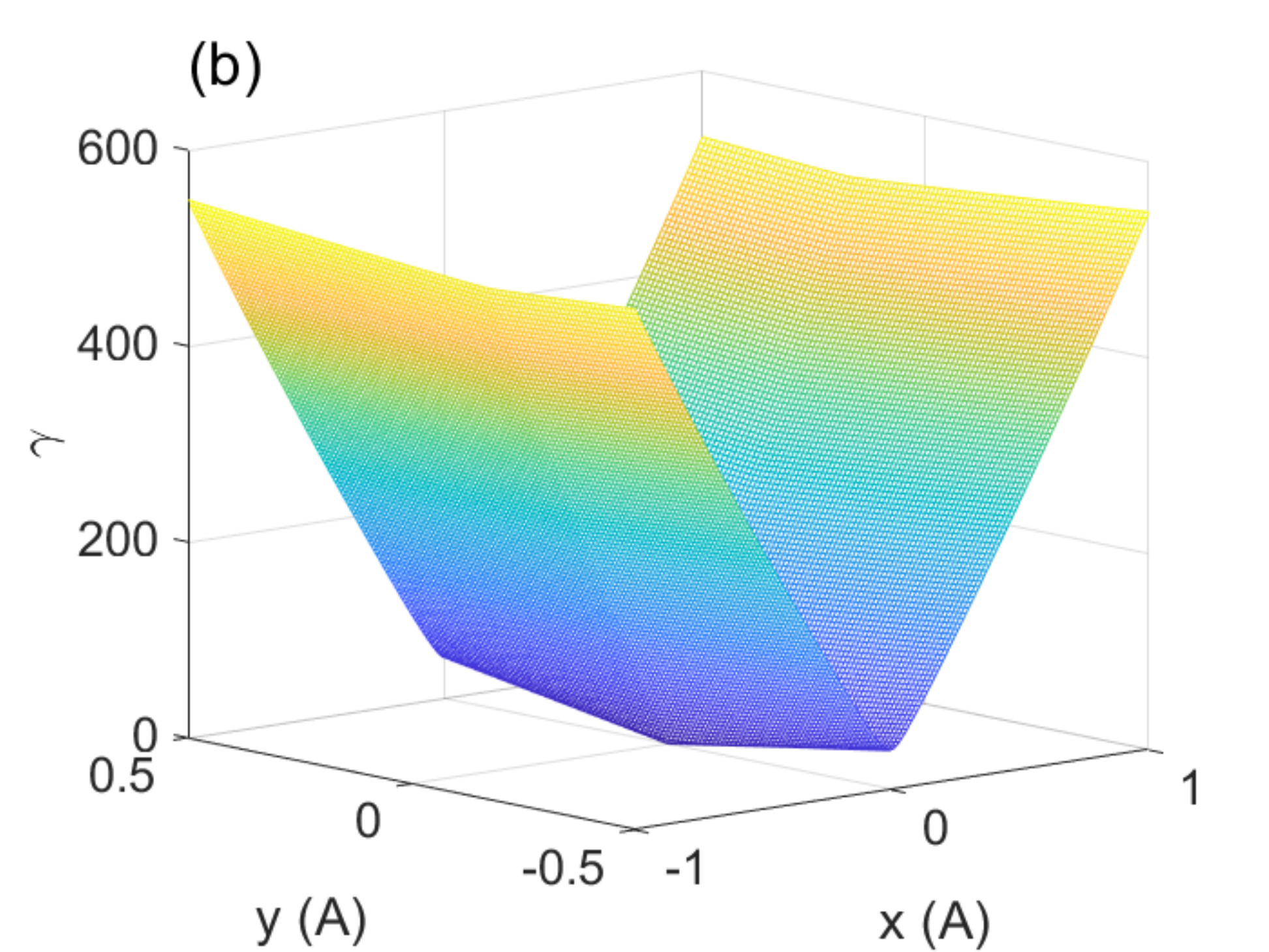}
    \caption{The rate function $\gamma$ for (a) the truncated potential $\tilde{V}$ and (b) the full double well potential $V$. $\sigma = 0.2$ is used to  parameterize Eq.\ref{delta}. 
    Note that the V-shaped profile near the bottom tip is similar for both cases, but the magnitude of the rate functions differs by several orders of magnitude and despite the rate of creating particles are different between two potentials, the relative rate of particle creation in a given simulation remains similar.
    }
    \label{gammas}
\end{figure*}

Here, we take advantage of the fact that dynamics in many important chemical scenarios, e.g., ground-state dynamics at ambient temperatures, stay relatively low in energy and remain quite local in the sense that the simulation grid is typically truncated a certain small distance (nanometers) away from the chemical system of interest. Since for most potentials that are local and bounded, singularities do not exist \cite{muswag16} we will truncate the polynomial potentials in our simulations beyond energies that we do not expect our simulations to reach. In order to make the Wigner transform of the truncated potential simple, we train a neural network to fit the truncated potential as a sum of Gaussian functions. By the Cybenko's theorem, the fitting Gaussian functions can be arbitrarily close to the fitted truncated potential\cite{mlmit,dl}. By taking the Fourier transform of the fitted Gaussian functions, we get a smooth and bounded $V_w$ that does not have any singularities. Importantly, since the functional form is known analytically the burden of storing the Wigner potential numerically on a grid is lifted as well. Removing the need to store the Wigner potential on a grid is an important step towards making higher dimensional simulations of the Wigner equation computationally feasible.

To fit the truncated potential, we used a Multi Layer Perceptron (MLP) model with a single hidden layer and Gaussian activations to fit the double well potential typically used to model a proton transfer reaction.\cite{makri87, rom91} The functional form for the full double well potential is given by
 \begin{equation}\label{potential}
	V(x, y) = -\frac{1}{2} a_0 x^2 + \frac{1}{4} c_0 x^4 + \frac{1}{2} m\omega^2 y^2 - cxy,
\end{equation}

\noindent
where $m = 1837$ Da, $\omega = 2980$ cm$^{-1}$, and $c = 2.34$ eV/\AA$^2$. The potential along the $x$-direction is a symmetric double well, where the distance between the minima and the local maximum is $x_r = \sqrt{a_0/c_0} = 0.502$ \AA, and the height of the barrier is $E_a = a_0^2/4c_0 = 0.27$ eV. The truncated potential is generated by
\begin{equation}\label{potentrunc}
    \hat{V} = \min(V, V_{thr})
\end{equation}

\noindent
where we set the truncation threshold $V_{thr}$ at 4eV with the postulate that the details of the potential beyond this energy scale will not matter for low energy dynamics of nuclei. The fitted potential has the form
\begin{equation}\label{nnmodel}
    \tilde{V} = \sum_i A_i e^{-a_i x^2 - b_i xy + c_i y^2} + h_i,
\end{equation}

\noindent
where all symbols with subscript $i$ are parameters to be optimized. In this work the index $i$ ran from merely from $1$ to $3$ and back-propagation algorithm was used for the gradient decent optimization\cite{dl}. The fitted potential is then used to propagate dynamics. The truncated potential and the MLP fit are shown in Fig.\ \ref{fitted}. We can see an excellent agreement between the MLP fitted potential $\tilde{V}$ and the truncated potential $\hat{V}$ at the bottom of the wells, and an acceptable disagreement between the two beyond the truncation threshold. Hereafter, the term truncated potential will refer to the Gaussian MLP fit $\tilde{V}$.

So far we operated under that assumption that truncating a potential at sufficiently high energy will lead to no significant deviations in the generated dynamics. In order to explore what may be different we need to compare the auxiliary functions $\gamma$ that determine the rate of particle creation for the two cases: $\gamma$'s for the full $V$ and the truncated double-well $\tilde{V}$ potentials are shown in Fig.\ \ref{gammas}. Both $\gamma$'s shows V-shaped profile of the absolute value function along the coordinate axes near the tip while the heights of the two functions differ by two orders of magnitude. We therefore expect the relative rate of scattering throughout the phase space to be approximately retained while the rate at which signed particles accumulate in the simulation to differ drastically. 

\section{results and discussion}\label{a}

\begin{figure*}
    \includegraphics[width=0.8\paperwidth]{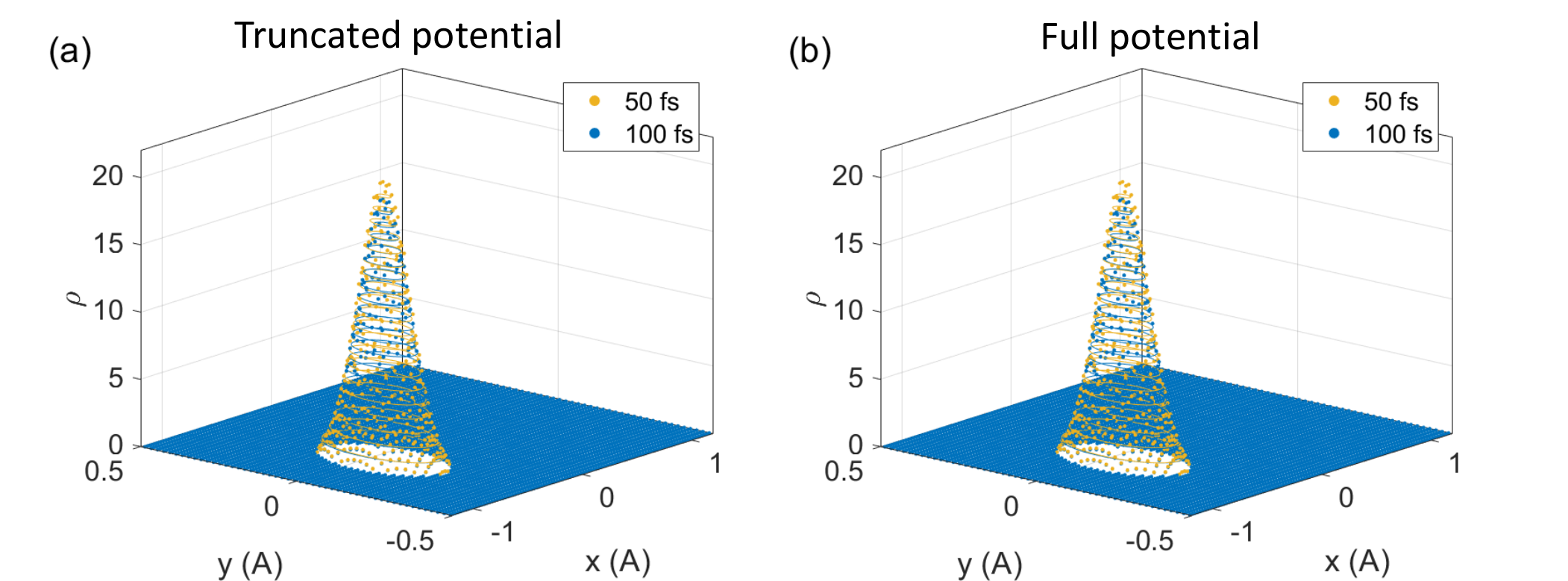}
    \caption{Benchmarking the Wigner quasi-probability function dynamics generated using SPMC (dots) against exact solutions (contours) for (a) the truncated potential $\hat{V}$ and (b) the full harmonic potential $V$. The computational advantage of our truncation/hyperparameterization scheme is demonstrated by the fact that similar dynamics are generated in (a) and (b) but in (b) the time step is 0.01 fs, and the rate of annihilating signed particles is once every 10 steps while in (a) the time step is 10 times smaller: 0.001 fs and the rate of annihilating signed particles is once every five steps.}
    \label{fw}
\end{figure*}

To demonstrate the benefits of our approach we apply our method to simulate the phase-space dynamics of a proton in a two-dimensional double well potential. The potential $V$ for this model is given by Eq.\ \eqref{potential} and the Wigner potential $V_w$ is then
\begin{multline}\label{vwd}
	V_w(x, y, p_x, p_y) = \left( -a_0 x + c_0 x^3 - cy \right) \delta'(-p_x) \\
	- \frac{1}{4} c_0 \hbar^2 x \delta'''(-p_x) + \left( m\omega^2 y - cx \right) \delta'(-p_y).
\end{multline}

\noindent
The derivatives of the delta functions give rise to the aforementioned singularities. In our numerical handing of this function we resort to the Gaussian representation of a delta function and then evaluate the $V_w$ and the $\gamma(x, y)$ functions approximately by choosing a sufficiently small width $\sigma$:
\begin{equation}\label{delta}
    \delta(x) = \lim_{\sigma \to 0^+} \frac{1}{\sigma\sqrt{\pi}}\
    e^{-x^2 / \sigma^2}.
\end{equation}

\noindent
The Wigner potential for the truncated potential $\tilde{V}$ is obtained by a Fourier transform of the Gaussian MLP model Eq.\ \eqref{nnmodel}
\begin{equation}\label{Vw_MLP}
    \tilde{V}_w = \sum_i \frac{\sin(2p_x x + 2p_y y)}{\pi\hbar^3\sqrt{\frac{a_i c_i}{4} - \frac{b_i^2}{16}}}\
    \exp \left( \frac{4a_i p_y^2 + 4c_i p_x^2 - 4b_i p_x p_y}{\hbar^2 (b_i^2 - 4a_i c_i)} \right), \\
\end{equation}

\noindent
where the symbols with subscript $i$ correspond to the neural network parameters in Eq.\ \eqref{nnmodel}.

We now have all the ingredients we need to generate the phase space dynamics of a proton in a double well using the SPMC algorithm. We will first run short (100fs) simulations to benchmark our SPMC protocol against exact solutions for $V$ and for $\tilde{V}$ and then proceed to run longer trajectories to demonstrate numerical stability. We first run the dynamics of propagating the wavepacket on the truncated potential and the full double-well potential. The initial state is a Gaussian that is centered at the minimum of the left well and corresponds to the lowest eigenstate of the locally harmonic potential of the full potential, where the frequency of the approximate harmonic potential at the bottom in the $x$-direction can be expressed as $\omega_x = \sqrt{2a_0^2/m}$. The initial wavepacket has the form
\begin{equation}
	\psi_0(x, y) = \sqrt{\frac{m\sqrt{\omega_x \omega}}{\pi\hbar}}\ e^{\frac{-m\omega_x}{2\hbar} (x + x_r)^2}
	e^{\frac{-m\omega}{2\hbar} y^2}.
\end{equation}

\noindent
The Wigner function that corresponds to the initial state is
\begin{equation}
	f_{w, 0}(x, y) = \frac{1}{(\pi\hbar)^2} e^{-m\omega_x (x + x_r)^2 / \hbar}
	e^{-p_x^2 / \hbar m\omega_x} e^{-m\omega y^2 / \hbar} e^{-p_y^2 / \hbar m\omega}
\end{equation}
and about 500,000 signed particles were used to generate the initial state in the SPMC simulations.
\begin{figure*}
    \includegraphics[width=0.8\paperwidth]{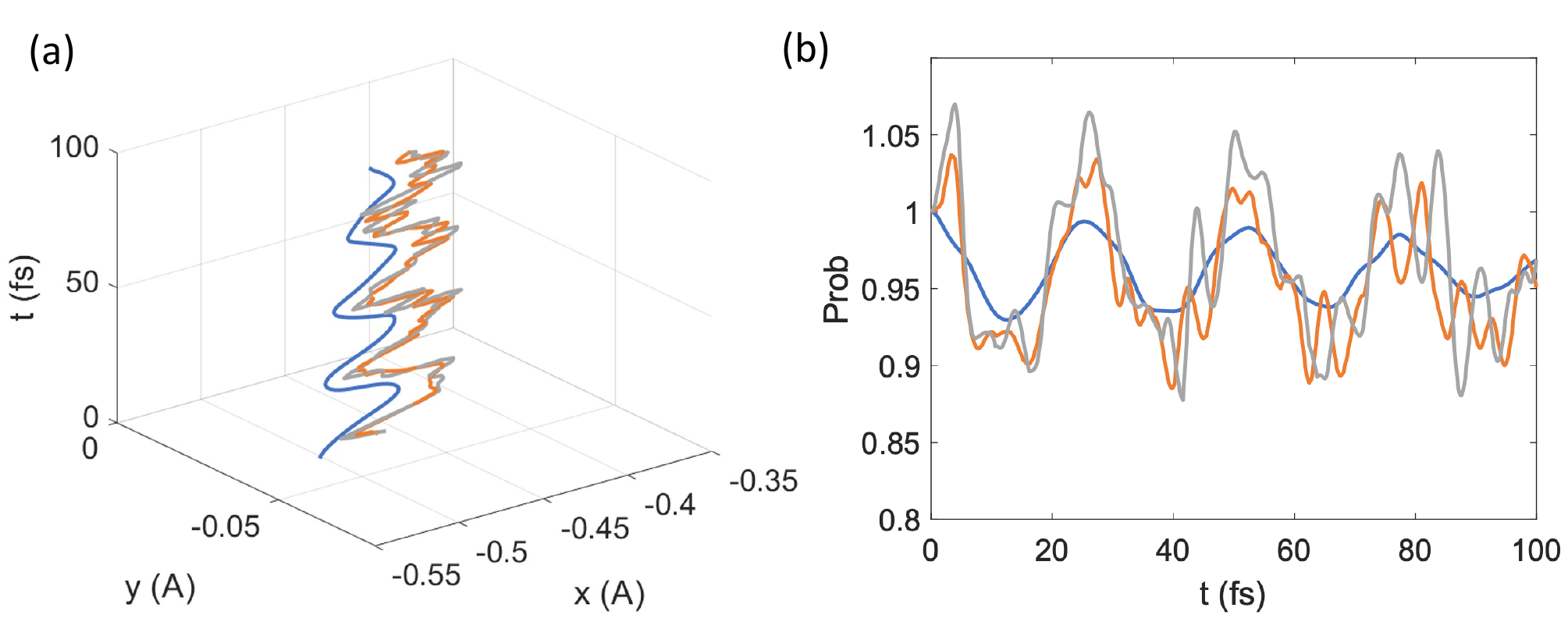}
    \caption{Benchmarking dynamics of observables generated using SPMC against the exact solution (blue) for the truncated potential $\tilde{V}$ (gray) and the full potential $V$ (orange). The expectation value of the position operator as a function of time is shown in (a). In (b) the survival probability of the wavepacket is tracked. The main oscillating frequency in the simulations match, the noisy fluctuations in the SPMC results are caused by the numerical integration protocol: random fluctuations in the reconstruction of the Wigner function leads to minor numerical artifacts. Furthermore, a coarser grid is used in SPMC than in the exact solution and this accounts to a large extent for the of additional oscillations in the SPMC results.}
    \label{x}
\end{figure*}

The auxiliary function $\gamma$ for $\tilde{V}$ is calculated by numerically integrating the Wigner potential $\tilde{V}_w^+$ of Eq.\ \eqref{Vw_MLP}. For the full potential $V$, $\gamma$ is calculated by integrating $V_w^+$ numerically using Eq.\ \eqref{delta} with $\sigma = 0.2$. In both cases, we propagate the system for 100 fs and compare the outcomes to the exact solution. Exact solutions are obtained by applying the finite difference method to the time-dependent Schr\"{o}dinger equation. 

In Fig.\ \ref{fw}(a) and (b), the Monte Carlo simulation shows a good correspondence with the exact solution in the density plots for both simulations along the duration of the trajectory. The two simulations are very close to each other as well, indicating that truncation does not manifest itself in low energy dynamics. The rate of creating of new particles in the $\tilde{V}$ simulation is at 0.003\% per time step while the time step needs to be kept small 0.001fs, and the annihilation frequency is set to every 5 steps. In the full potential $V$ simulation the rate of creating new particles is much higher --- at 2\% per time step, but the time step is larger 0.01fs and the annihilation frequency is lower (every 10 steps) compared to the truncated potential. The simulation hyperparameters for the truncated potential $\tilde{V}$ were found by a systematic grid search of time step and annihilation step frequency. We note that the larger time-step in the full potential was found almost by accident as a small and disconnected 'island' of stability in the parameter space. A deeper insight into why such hyperparameters exist, why they are separated from a much larger set of stable hyperparameters which happen to be limited to much shorter time steps, and how to find them is left to future work. 

The dynamics in these potentials are very mild and they are difficult to observe from the Wigner function directly. We therefore present the dynamics of the expectation values for the position and the survival probability dynamics in Fig.\ \ref{x}(a) and (b) respectively.\cite{light} The survival probability is defined as:
\begin{equation}
    \mathbf{P} = |\langle \psi_0 | \psi_t \rangle|^2
    = (2\pi\hbar)^n \iint f_w(0) f_w(t)\ d^n\mathbf{x} d^n\mathbf{p}.
\end{equation}

\noindent
Comparing our results with the exact solution and also with those in the literature \cite{makri87} we see that there are extra frequency components for Monte Carlo results while the main oscillating frequency matches. The extra frequencies appear because a course grid was used to reconstruct the Wigner function from signed particles and also because of the re-sampling process. Numerically integrating on a course grid would lead to numerical errors, and the situation gets worse because calculating the survival probability implies integrating with twice as many dimensions. Other than numerical artifacts from the integration protocol, the reconstructed Wigner function possesses random fluctuations. Fluctuations are not visible in Fig.\ \ref{fw} because real space densities are marginal distributions of the Wigner functions. To calculate the densities, we integrated two momentum dimensions of the Wigner function in which four dimensions are in total and this integration eradicated fluctuations. In Fig.\ \ref{x}(b), two Monte Carlo results match quite well despite deviating from the exact solutions. We conclude that the fine movements of the wavepacket cannot be captured with high resolution by the Monte Carlo method and it should be restricted to estimating only the dominant behavior with the benefit of a much lower cost than the exact calculation.

\begin{figure*}
    \includegraphics[width=0.8\paperwidth]{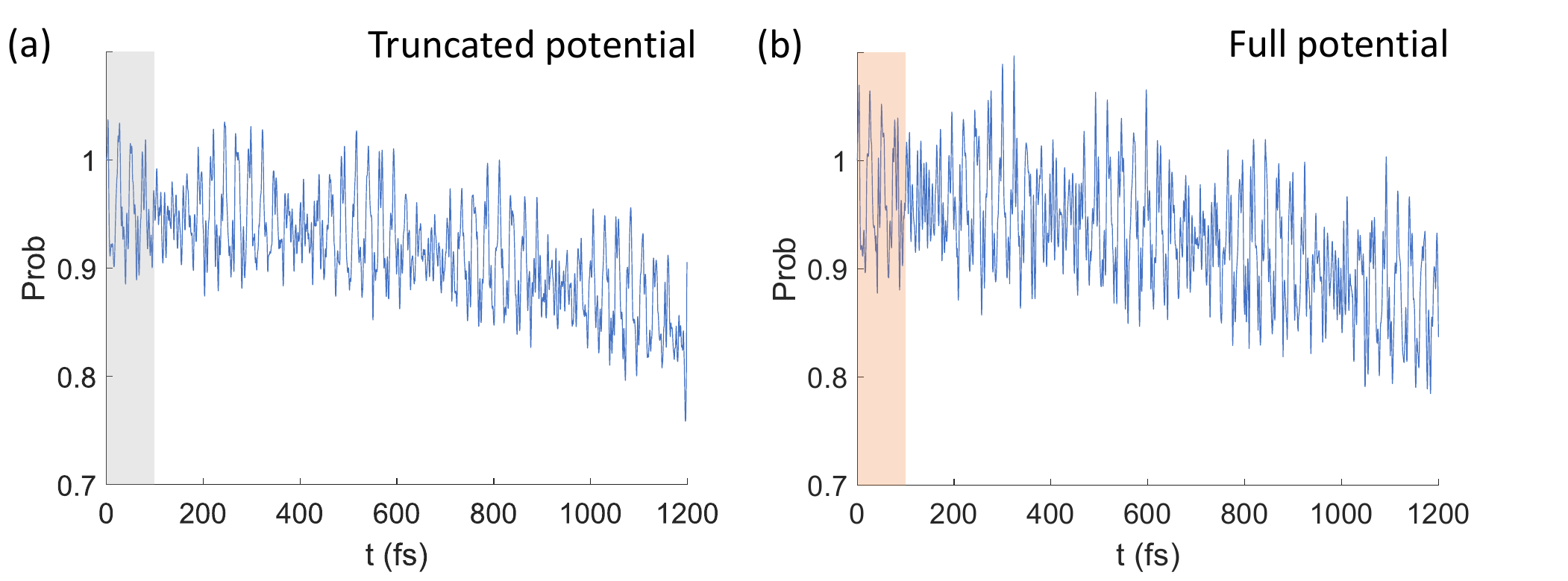}
    \caption{Stability of the SPMC simulation: the survival probability in long-time (picoseconds) trajectories is shown for (a) the truncated potential $\tilde{V}$ and for (b) the full potential $V$. The color stripe highlights the 100fs trajectory that we directly compare to the exact solution, see Fig. 4(b), and the color of the stripe corresponds to the color of the lines in Fig.\ \ref{x}(b). The decay of the probability indicates that part of the wavepacket has tunneled to the right well.}
    \label{survival}
\end{figure*}

Finally, in order to demonstrate feasibility of longer time-scale simulations we have generated stable 1.5ps trajectories which ran on an ordinary desktop computer and took 350 cpu hours to complete. The calculated survival probability dynamics for the simulations generated by the truncated and the full potentials are shown in Fig. \ref{survival}. We notice no significant differences in the features of the dynamics generated by the two potentials pointing at the fact that truncation did not negatively affect the physics of the simulation. Furthermore, we compare the dynamics to a similar albeit not identical simulation in Ref. \cite{makri87} and observe that the oscillation patterns are quite similar. 

\section{conclusions}

To conclude, we have made a step forward in the direction of performing long-time phase-space dynamics in high-dimensional molecular systems by solving some of the difficulties that arise when solving the Wigner equation numerically using the Signed Particles Monte Carlo approach. We have encoded a model potential in a neural network in order to avoid the computational cost of storing a high-dimensional object, the Wigner potential, on a grid alleviating a cost that makes phase-space molecular dynamics infeasible for most systems and we have improved further the SPMC algorithms. We have demonstrated numerical stability of our implementation using a potential that models a proton-transfer reaction. The work towards including dissipation and towards N-dimensional simulations of molecular dynamics in phase-space is underway.

\begin{acknowledgments}
We acknowledge the support of the Natural Sciences and Engineering Research Council of Canada (NSERC). Nous remercions le Conseil de recherches en sciences naturelles et en génie du Canada (CRSNG) de son soutien.
\end{acknowledgments}

\appendix

\section{Derivation of Eq.\ \eqref{deltas}}

In this section, we derive Eq.\ \eqref{deltas} and prove that it is an unbiased estimator to Eq.\ \eqref{eomdt}. The discretization of the Wigner function is done by
\begin{equation}\label{appa}
\begin{split}
    f_w(\eta) &= \frac{1}{n_+ - n_-} \sum_i \sgn \big( f_w(\eta_i) \big)
    \delta(\eta - \eta_i) \\
    &= \frac{1}{n_+ - n_-} \E \Bigg[ \sum_i
    \sgn \big( f_w(\eta_i) \big) \ind_S(\omega(\eta_i)) \Bigg],
\end{split}
\end{equation}

\noindent
where $\ind$ is the indicator function and $S$ is the support of $f_w$. $n_+$ and $n_-$ refers to
\begin{equation}
    n_\pm = \sum_i \sgn \big( f_w(\eta_i) \big),
    \quad \text{if} \; f_w(\eta_i) \gtrless 0.
\end{equation}

\noindent
$\omega$ is the sample point that is used to sample $f_w$, which is an alternative but more rigorous way of expressing the delta functions in the first equality in Eq.\ \eqref{appa}. $\omega(\eta_i)$ indicates that this sample point samples $f_w$ in the neighborhood of the phase space point $\eta_i$. The indicator function in Eq.\ \eqref{appa} indicates that if a sample point $\omega_i$ samples $f_w$ at a point where $f_w = 0$ everywhere in the neighborhood of $\eta_i$, then this is a meaningless sample. Following the discretization process, we express the Wigner function in terms of the random variable $F_w$
\begin{equation}
    F_w(\omega_i) = \frac{\sgn \big( f_w(\eta_i) \big)}{n_+ - n_-}
    \equiv u_i,
\end{equation}

\noindent
where $\omega_i = \omega(\eta_i)$ is the shorthand notation. $u_i$ represents the portion that the sample point $\omega_i$ samples the $f_w$.

If we use the Schr\"{o}dinger picture to describe the time evolution, we may express the time evolution as an operator $U$ acting on the initial Wigner function
\begin{equation}
    f_w(t) = U(t, 0) f_w(0) = U_t f_w(0),
\end{equation}

\noindent
and therefore
\begin{equation}
    F_{w, t}(\omega_t^i) = U_t F_w(\omega_0^i).
\end{equation}

\noindent
From Eq.\ \eqref{eomdt}, the time evolution operator satisfies the following relation:
\begin{equation}\label{appb}
    U_{t+dt} = (1 - \gamma dt) U_t + \gamma dt \int
    K(\eta,\ d\eta')\ U + o(dt).
\end{equation}

\noindent
where $K$ is the Markov kernel. This is the Kolmogorov's backward equation, and the process that is described by it is called the jump process.\cite{prob} For jump processes, the time evolution of the expectation of the sampling is described by the Dynkin's formula. However, a corollary from the Dynkin's formula is more useful for our purposes than the formula itself, which is\cite{wagner16}
\begin{equation}\label{appc}
    \frac{d}{dt} \sum_i F(\omega_i) \ind_S(\omega_i) = \sum_i \ind_S(\omega_i)
    \bigg( \sum_j \hat{K} F(\omega_j) \bigg).
\end{equation}

\noindent
The kernel $\hat{K}$ is the descrete version of the $K$ in Eq.\ \eqref{appb}, and describes the ``jump'' only but not the ``wait'' in the jump process. $F$ is an arbitrary test random variable, and it and its corresponding test function $f$ satisfies
\begin{equation}
    F(\omega(\eta_i)) = u_i f(\eta_i).
\end{equation}

\noindent
The test function $F$/$f$ can be designated as any quantum observable. If the reader finds it's difficult to read, one can always define $f$ as a constant unit function: $f(\eta) = 1,\; \forall \eta$. If the kernel $\hat{K}$ satisfies
\begin{multline}\label{appd}
    \sum_i \ind_S(\omega_i) \bigg( \sum_j \hat{K} F(\omega_j) \bigg)
    = \sum_i u_i \ind_S(\omega_i) \\
    \times \frac{1}{2} \int_{\Bbb{R}^n} V_w(\eta') \Big( f(\eta_i^\Delta)
    - f(\eta_t^\Sigma) \Big)\ d^n\mathbf{p}',
\end{multline}

\noindent
then the corollary \eqref{appc} becomes

\begin{widetext}

\begin{equation}\label{appe}
\begin{split}
    \frac{d}{dt} \sum_i u_i \ind_S(\omega_i)
    &= \sum_i u_i \ind_S(\omega_i)
    \frac{1}{2} \int_{\Bbb{R}^n} V_w(\eta')
    \times \Big( f(\eta_i^\Delta) - f(\eta_i^\Sigma) \Big)\ d^n\mathbf{p}' \\
    &= \sum_i \frac{1}{2} \int_{\Bbb{R}^n} V_w(\eta')
    \Big( u_i \ind_S(\omega_i) f(\eta_i^\Delta)
    - u_i \ind_S(\omega_i) f(\eta_i^\Sigma) \Big)\ d^n\mathbf{p}' \\
    &= \sum_i \frac{1}{2} \int_{\Bbb{R}^n} V_w(\eta')
    \Big( u_i \ind_S \big( \omega(\eta_i^\Delta) \big) f(\eta_i^\Delta)
    - u_i \ind_S \big( \omega(\eta_i^\Sigma) \big) f(\eta_i^\Sigma)
    \Big)\ d^n\mathbf{p}'
\end{split}
\end{equation}

\end{widetext}

\noindent
The last equatliy of Eq.\ \eqref{appe} holds because changing $\eta$ in the $\omega$ does not change the sampling weight. By letting $f$ to be the unit function and taking the expectation, Eq.\ \eqref{appe} becomes the Wigner equation of motion \eqref{eomt}:
\begin{equation}\label{appf}
	\frac{d f_w(\eta_t)}{dt}
	= \int_{\Bbb{R}^n} \frac{V_w(\eta_t')}{2}
	\Big( f_w(\eta_t^\Delta) - f_w(\eta_t^\Sigma) \Big)\ d^n\mathbf{p}'.
\end{equation}

To find the kernel $\hat{K}$, we inspect what happens between $t$ and $t + dt$ in the jump process. During this period of time, jump occurs to some of the states as characterized by the survival probability. The sampling to $f_w$ changes and is characterized by the last equality in Eq.\ \eqref{appe}:
\begin{multline}\label{appg}
    \frac{d}{dt} \sum_j u_j \ind_S(\omega_j) =
    \sum_j \gamma \int_{\Bbb{R}^n} \frac{V_w(\eta')}{2\gamma}
    \Big( u_j \ind_S \big( \omega(\eta_j^\Delta) \big) \\
    + (-u_j) \ind_S \big( \omega(\eta_j^\Sigma) \big) \Big)\ d^n\mathbf{p}'
\end{multline}

\noindent
where the summation index $j$ only runs through the points for which jump occurs. In Eq.\ \eqref{appg}, we multiply and devide $\gamma$ to normalize $V_w$. To separate $\hat{K}$ from $\omega$'s, we specify the point that is to be sampled in the indicator function $\ind_S \big( \omega(\eta_j^\Delta) \big) = \ind_{\{\eta^\Delta\}} (\omega_j)$. Therefore, from Eq.\ \eqref{appc} and \eqref{appg},
\begin{multline}
    \sum_j \hat{K} \omega_j = \sum_j \gamma \int_{\Bbb{R}^n} d^n\mathbf{p}'\ \frac{V_w(\eta')}{2\gamma} \\
    \times \Big( (-u_j) \ind_{\{\eta_j^\Delta\}}
    + (u_j) \ind_{\{\eta_j^\Sigma\}} \Big) (\omega_j),
\end{multline}

\noindent
Note that the reverse of the sign in the $u_j$'s. The sample points $\omega_j$'s at $t$ will become a sample point $\omega_i$ at $t + dt$. The sampling weight from the $\omega_j$'s is transferred to $\omega_i$. Therefore, we need to negate the weight from the $\omega_j$'s by reversing the sign. Thus, the kernel $\hat{K}$ writes
\begin{equation}
    \hat{K} = \gamma \int_{\Bbb{R}^n} d^n\mathbf{p}' \frac{V_w(\eta')}{2\gamma}
    \Big( -u_j \ind_{\{\eta_j^\Delta\}} + u_j \ind_{\{\eta_j^\Sigma\}} \Big).
\end{equation}

\noindent
However, the non-jumping states are not characterized by $\hat{K}$, although they do not contributes to the change in the sampling weight. Thus, we add the non-jumping states to be
\begin{multline}\label{apph}
    \hat{K} = \gamma \int_{\Bbb{R}^n} d^n\mathbf{p}'\
    \frac{V_w(\eta')}{2\gamma}
    \Big( -u_j \ind_{\{\eta_j^\Delta\}}
    + u_j \ind_{\{\eta_j^\Sigma\}} \Big) \\
    + \left( 1 - \frac{V_w(\eta')}{2\gamma} \right) \ind_{\emptyset}
\end{multline}

\noindent
where $\ind_\emptyset(\omega) = 0$ for all sample points. The kernel $\hat{K}$ is what we're seeking for in Eq.\ \eqref{appb} or \eqref{appc}. By reversing the deriving sequence, that is going from Eq.\ \eqref{apph} back to \eqref{appc}, it shows that the kernel $\hat{K}$ is a unbiased estimator to the jump process. The sample point $\omega$, along with the sign $u$, is what we called the signed particles in the main text. By substituting $\hat{K}$ back into the Kolmogorov's backward equation \eqref{appb}, taking expectation to the sample points, and replacing the $\omega$'s with the delta functions, we get the propagator in Eq.\ \eqref{deltas}.

\nocite{*}
\bibliography{doc}

\end{document}